\documentclass[journal]{IEEEtran}

\usepackage{cite}
\usepackage{amsmath,amssymb,amsthm}
\usepackage{graphicx}
\usepackage{booktabs}
\usepackage{xcolor}
\usepackage{algorithm}
\usepackage{algpseudocode}
\usepackage[hidelinks]{hyperref}

\graphicspath{{figures/}}

\newtheorem{theorem}{Theorem}
\newtheorem{lemma}{Lemma}

\newtheorem{definition}{Definition}
\newtheorem{problem}{Problem}

\newcommand{\mis}{\ensuremath{\operatorname{MIS}}}
\newcommand{\valpha}{\ensuremath{\boldsymbol{\alpha}}}
\newcommand{\rvalpha}{\ensuremath{\tilde{\boldsymbol{\alpha}}}}
\newcommand{\cross}{\textsf{CROSS}}
\newcommand{\crossedge}{\textsf{CROSS+EDGE}}
\newcommand{\branchcopy}{\textsf{BRANCH-COPY}}

\algrenewcommand\algorithmicrequire{\textbf{Input:}}
\algrenewcommand\algorithmicensure{\textbf{Output:}}

\hypersetup{pdftitle={Encoding Maximum Independent Set on Triangular Rydberg Atom Arrays with Uniform Detuning},pdfauthor={Jinguo Liu, Xiwei Pan, Xian Wu, and Chunfan Yao}}

\begin{document}

\title{Encoding Maximum Independent Set on Triangular Rydberg Atom Arrays with Uniform Detuning}

\author{Jinguo~Liu, Xiwei~Pan, Xian~Wu, and Chunfan~Yao%
\thanks{This work was partially supported by the National Key R\&D Program
of China under Grant No.~2024YFB4504004 and the National Natural Science
Foundation of China under Grant No.~12404568.
The authors are listed in alphabetical order by surname.
\textit{(Corresponding author: Xiwei Pan.)}}%
\thanks{Jinguo Liu and Xiwei Pan are with the Thrust of Advanced Materials,
The Hong Kong University of Science and Technology (Guangzhou), Guangzhou, Guangdong,
China (e-mail: xpan432@connect.hkust-gz.edu.cn).}%
\thanks{Xian Wu and Chunfan Yao are with the Thrust of Artificial
Intelligence, Information Hub, The Hong Kong University of Science and
Technology (Guangzhou), Guangzhou, Guangdong, China.}%
\thanks{Jinguo Liu and Xian Wu are also with QudeLeap Research,
Shanghai 200030, China.}}

\markboth{Liu et al.: Encoding Maximum Independent Set on Triangular Rydberg Atom Arrays}%
{Liu \MakeLowercase{\textit{et al.}}: Encoding Maximum Independent Set on
Triangular Rydberg Atom Arrays with Uniform Detuning}

\maketitle

\begin{abstract}
Quantum optimization aims to solve hard combinatorial problems through
controlled quantum dynamics. Using Rydberg atom arrays on a King's
lattice, Ebadi \emph{et al.}\ [\emph{Science} \textbf{376}, 1209 (2022)]
observed a superlinear quantum speedup for maximum independent set
(MIS), but relied on classical postprocessing to repair constraint
violations. Whether a quantum device without these violations can
achieve better performance scaling remains unknown. Because triangular lattices better separate intended from unwanted
atomic interactions, we use AI-assisted gadget design to compile MIS on
arbitrary $n$-vertex graphs onto triangular arrays with uniform detuning
and exact solution recovery. One construction guarantees $O(n^2)$ atoms,
and a second yields smaller layouts on tested sparse graphs. In coherent
quantum annealing simulations of the nonplanar graph $K_{3,3}$, longer
annealing rapidly suppresses violations on the triangular layout and
steadily raises the MIS success probability without postprocessing. On
the King's layout, however, violations persist and the success
probability fails to grow even with postprocessing. This scaling
contrast provides evidence that annealing without postprocessing can be
much more efficient.
\end{abstract}

\begin{IEEEkeywords}
Quantum compilation, Rydberg atom arrays, maximum independent set,
placement and routing.
\end{IEEEkeywords}

\section{Introduction}
\label{sec:intro}

\begin{figure*}[t]
\centering
\includegraphics[width=\textwidth]{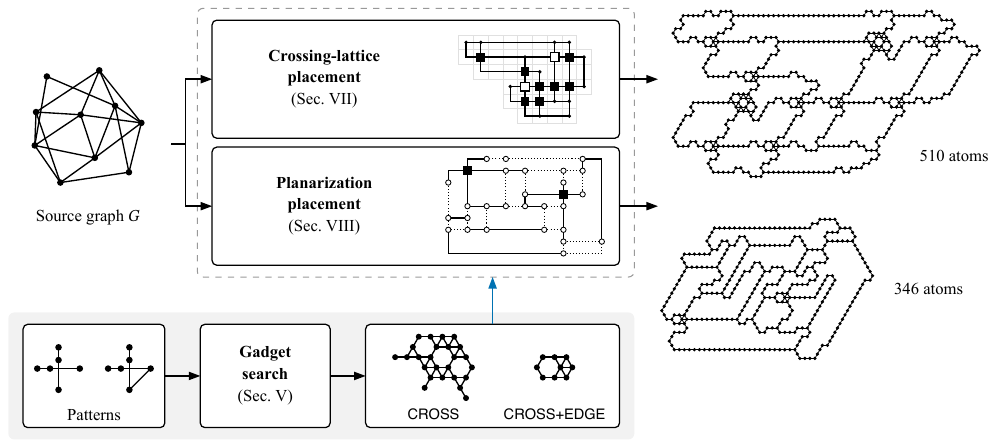}
\caption{\label{fig:flow}%
Compilation of a nonplanar 10-vertex graph. Gadget search
(Sec.~\ref{sec:search}) builds the reusable crossing gadgets. For each
source graph, the crossing-lattice and planarization placements construct
atom layouts from these gadgets. They use 510 and 346 atoms, respectively,
shown at the same scale. Lattice neighbors define the blockade edges.}
\end{figure*}

\IEEEPARstart{A}{rrays} of neutral atoms encode one class of
maximum-independent-set (MIS) problems natively. Optical tweezers hold
the atoms at programmable positions, and a laser drives each atom toward a
highly excited Rydberg state~\cite{browaeys2020many}. In the ideal
blockade limit, two atoms closer than the blockade radius cannot both be
excited. So the ground state of a suitably driven array is a maximum
independent set of its blockade graph, which connects atoms within the
blockade radius~\cite{pichler2018quantum,ebadi2022quantum}. Using this
encoding, Ebadi \emph{et al.}~\cite{ebadi2022quantum} ran quantum
optimization on King's-lattice arrays of up to 289 atoms. On the hardest
instances, they observed a superlinear quantum speedup over simulated
annealing in finding an exact MIS. This was the first such evidence for
Rydberg-atom quantum optimization.

These experiments also point to a clear opportunity for further
gains. In the published data for their 179-atom showcase graph, more
than 90\% of raw configurations violate an independence constraint, and
about $0.03\%$ form maximum independent sets as
measured~\cite{ebadi2022quantum}. Ebadi \emph{et al.}\ attribute part of
these violations to the finite interaction energy between nearby atoms,
or imperfect blockade, and apply classical postprocessing to every shot:
\emph{vertex reduction} removes atoms involved in violations, and
\emph{vertex addition} greedily adds atoms until the set is maximal.
Even with both steps, the MIS probability on this graph remains below
1\%. Because postprocessing operates only after measurement, a natural
question is whether a quantum device that suppresses these violations
natively can achieve better performance scaling.

The lattice geometry governs how strongly unwanted interactions compete
with the blockade. On the King's lattice with spacing $a$, the blockade
radius must lie between $\sqrt2a$ and $2a$, so the interaction strengths
at these distances differ by a factor of $(2/\sqrt2)^6=8$. On a
triangular lattice, the corresponding distances are $a$ and $\sqrt3a$,
giving a factor of $27$, more than three times larger. Pan \emph{et
al.}~\cite{pan2025encoding} used this geometry and found
substantially fewer blockade violations on triangular arrays in
annealing simulations. Their encoding uses vertex weights from 1 to 4,
where a selected vertex contributes its weight to the objective, which
requires site-resolved detuning control and
calibration~\cite{de2025demonstration}.

On a triangular lattice whose blockade radius lies between the first and
second neighbor distances, the native problems are MIS on induced
subgraphs of the lattice. MIS remains NP-hard on this
class~\cite{bringmann2017maximum}, while general applications require
arbitrary graph connectivity. A compiler must therefore turn an
\emph{arbitrary} graph $G$ into a graph realizable by the array and map
its optimal solutions back. It produces atom positions whose blockade
graph is $H$, and an integer $c(G)$ such that $\mis(H)=\mis(G)+c(G)$,
where $\mis(G)$ is the size of a maximum independent set of $G$. A
decoding rule turns every maximum independent set of $H$ into one of
$G$. Such problem transformations are called \emph{reductions}. Every
problem in NP can be reduced to MIS in polynomial
time~\cite{karp1972reducibility}, so such a compiler brings every
problem in NP to the array. The task resembles the physical design of an
integrated circuit. Small sets of atoms with verified logical behavior,
called \emph{gadgets}, are placed on the lattice and joined by chains of
atoms, and the atom count plays the role of layout area. Existing
compilers achieve quadratic atom counts on both lattices. Nguyen
\emph{et al.}~\cite{nguyen2023quantum} and Pan \emph{et
al.}~\cite{pan2025encoding} compile arbitrary graphs onto the King's and
triangular lattices, respectively, using weighted gadgets. Liu \emph{et
al.}~\cite{liu2024unweighted} give an \emph{unweighted} encoding on the
King's lattice, in which every atom has weight one, so all atoms follow
a single global detuning schedule. Combining the stronger interaction
separation of the triangular lattice with uniform global detuning
therefore calls for an unweighted triangular-lattice encoding.

In this work, we introduce an unweighted, uniform-detuning compiler for
MIS on arbitrary graphs using triangular Rydberg atom arrays.
AI-assisted Boolean satisfiability (SAT) search supplies the crossing
gadgets (Sec.~\ref{sec:search}). We provide two complementary placement
methods: the crossing-lattice method guarantees at most $21.5n^2$ atoms
for any $n$-vertex graph (Sec.~\ref{sec:grid}), while the planarization
method exploits graph structure to produce smaller layouts on sparse
graphs (Sec.~\ref{sec:planar}). We establish an unconditional
$\Omega(n^2)$ lower bound for chain-and-gadget layouts
(Sec.~\ref{sec:lower}) and show that the quadratic atom-count exponent
is conditionally optimal for general deterministic polynomial-time exact
compilation (Sec.~\ref{sec:discussion}). Benchmarks compare the atom
counts of both placements with unweighted King's-lattice encodings
(Sec.~\ref{sec:scaling}). In coherent quantum annealing simulations of
the nonplanar graph $K_{3,3}$ (Sec.~\ref{sec:gap}), longer sweeps
rapidly suppress blockade violations on the triangular array and
steadily raise the direct MIS success probability without
postprocessing, whereas on the King's array violations persist and the
success probability fails to grow even with the postprocessing of Ebadi
\emph{et al.}

Section~\ref{sec:prelim} reviews the background and related work.
Section~\ref{sec:formulation} formulates the compilation problem and
outlines the compiler. Section~\ref{sec:composition} introduces the
gadgets, Sec.~\ref{sec:search} describes the gadget search,
Sec.~\ref{sec:lower} proves the quadratic lower bound,
Secs.~\ref{sec:grid}--\ref{sec:planar} present the placement algorithms,
Sec.~\ref{sec:results} reports the experimental results,
Sec.~\ref{sec:discussion} discusses optimality, experimental
implementation, and broader applications, and Sec.~\ref{sec:conclusion}
concludes.

\section{Preliminaries and Related Work}
\label{sec:prelim}

\subsection{MIS on Rydberg Atom Arrays}
\label{sec:rydberg}
The maximum independent set (MIS) problem asks for the largest set of vertices of
a graph $G=(V,E)$ in which no two are adjacent. We write $\mis(G)$ for its
size. It is NP-hard in general~\cite{karp1972reducibility} and remains so
on unit-disk graphs (UDGs)~\cite{clark1990unit}, whose vertices are connected only
within a fixed distance. It also remains NP-hard on the induced subgraphs
of the triangular lattice~\cite{bringmann2017maximum}.

A Rydberg atom array consists of neutral atoms held at programmable
positions by optical tweezers~\cite{manetsch2024tweezer}. A laser drives
transitions between each atom's ground state and a highly excited Rydberg
state. The Hamiltonian of the array is
\begin{equation}
 \mathcal H=\frac{\Omega}{2}\sum_i\sigma^x_i-\Delta\sum_in_i
 +\sum_{i<j}\frac{C_6}{|\mathbf x_i-\mathbf x_j|^6}\,n_in_j ,
 \label{eq:rydberg}
\end{equation}
where $\sigma^x_i$ couples the two states and $n_i$ projects onto the
Rydberg state. The laser parameters are the Rabi frequency $\Omega$ and
the detuning $\Delta$. Here $\mathbf x_i$ is the position of atom $i$, and
$C_6$ is the interaction
coefficient~\cite{jaksch2000fast,saffman2010quantum,browaeys2020many}.
The third term adds a strong, distance-dependent energy penalty when two
nearby atoms are both in the Rydberg state. This penalty produces the
Rydberg blockade. Our reduction uses the hard-blockade approximation:
simultaneous excitations within $R_b$ are forbidden, and interactions
beyond $R_b$ are neglected. At $\Omega=0$ and uniform $\Delta>0$, an
allowed configuration $I$ has energy $-\Delta|I|$, so its ground states
are exactly the maximum independent sets of the blockade graph.
In the full Hamiltonian~\eqref{eq:rydberg}, long-range $C_6/r^6$ tails
and finite blockade energy perturb the low-energy spectrum; our
annealing simulations in Sec.~\ref{sec:gap} retain all pairwise
interactions explicitly.

\begin{figure}[t]
\centering
\includegraphics[width=\columnwidth]{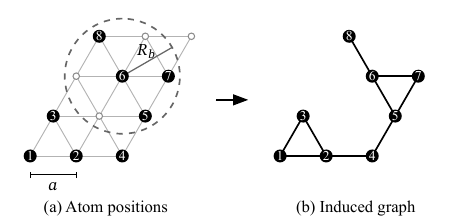}
\caption{\label{fig:lattice}%
(a) Atoms (filled circles) on a triangular lattice of spacing $a$, with
blockade radius $a<R_b<\sqrt3a$ (dashed circle).
(b) Their induced graph includes every lattice edge between atoms.}
\end{figure}

In this approximation, the array encodes MIS on induced
subgraphs of the lattice. Blockade prevents nearby atoms from being excited
together, and the laser detuning favors the largest allowed set of
excitations. Let the atoms sit
on a triangular lattice with spacing $a$, and let the blockade radius $R_b$
satisfy $a<R_b<\sqrt3a$. Two atoms then share a blockade edge exactly when
they are lattice neighbors, so the blockade graph is the subgraph of the lattice
induced by the atom positions [Fig.~\ref{fig:lattice}]. This paper targets unweighted MIS on such
graphs with uniform detuning. The
compiler chooses the atom positions so that an MIS of an arbitrary
source graph can be recovered from the ground state of the array, and it
minimizes the atom count.

\subsection{MIS Gadgets and Reduced \texorpdfstring{$\alpha$}{alpha}-Tensors}
\label{sec:tensor}

We adopt the graph-rewriting framework of Liu \emph{et
al.}~\cite{liu2024unweighted}, an unpublished manuscript, and include a
self-contained proof of its gadget criterion below.
An independent set is a set of vertices containing no adjacent pair; an
MIS is one of largest cardinality. We write $\mis(G)$ for this cardinality
in a graph $G$.

\begin{figure}[t]
\centering
\includegraphics[width=\columnwidth]{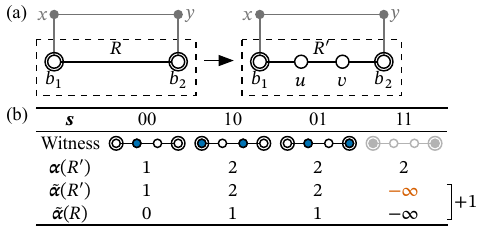}
\caption{\label{fig:tensor}%
(a) Rewrite $R\to R'$ preserving boundary vertices (double circles) and
exterior edges (gray).
(b) Largest independent sets for each boundary configuration (filled
vertices).
Dropping $11$ gives reduced tensors differing by $c=1$.}
\end{figure}

A rewrite replaces an induced subgraph $R$ of a host $G$ by $R'$
[Fig.~\ref{fig:tensor}(a)]. Both share an ordered \emph{boundary}
$\partial R=\partial R'=(b_1,\ldots,b_k)$, the only vertices allowed to
have neighbors outside $R$. We call the boundary vertices \emph{pins}.
The rewrite retains these vertices and their
exterior edges, replacing the interior and the edges within $R$.
We write $G[R\to R']$ for the resulting graph and $E=V(G)\setminus V(R)$
for the unchanged exterior.

\begin{definition}[MIS gadget]
$R\to R'$ is an MIS gadget if, for every host $G$, each MIS of
$G[R\to R']$ agrees on $E$ with some MIS of $G$.
\end{definition}

To check a rewrite locally, we record the best independent-set size for
each \emph{boundary configuration} $\mathbf s\in\{0,1\}^k$, where
$s_i=1$ means that $b_i$ is selected and $s_i=0$ means it is excluded.

\begin{definition}[$\alpha$-tensor]
The entry $\valpha(R)_{\mathbf s}$ is the largest independent-set size
in $R$ with boundary configuration $\mathbf s$, counting boundary
vertices. It is $-\infty$ if $\mathbf s$ is \emph{infeasible}, that is, if
two selected boundary vertices are adjacent.
\end{definition}

Figure~\ref{fig:tensor}(b) shows the four entries for a two-vertex boundary. For Boolean configurations $\mathbf s,\mathbf t\in\{0,1\}^k$, write $\mathbf s\preceq\mathbf t$
if $s_i\le t_i$ for every $i$, and $\mathbf s\prec\mathbf t$ if, in
addition, $\mathbf s\ne\mathbf t$.
For example, $10\prec11$, but $10\not\prec01$.

\begin{definition}[Reduced $\alpha$-tensor]
The reduced tensor $\rvalpha(R)$ has entries
\begin{equation}
 \rvalpha(R)_{\mathbf t}=
 \begin{cases}
  \valpha(R)_{\mathbf t}, &
   \valpha(R)_{\mathbf t}>\max_{\mathbf s\prec\mathbf t}\valpha(R)_{\mathbf s},\\
  -\infty, & \text{otherwise},
 \end{cases}
 \label{eq:reduced}
\end{equation}
where the maximum over an empty set is $-\infty$. Thus an entry is
retained exactly when it is finite and strictly larger than every entry
that selects fewer boundary vertices.
\end{definition}

Reduction drops a configuration when selecting fewer boundary vertices
does at least as well. In Fig.~\ref{fig:tensor}(b), $10$ and $11$ both
give size two, so $11$ is dropped.

\begin{theorem}[\cite{liu2024unweighted}]
\label{thm:reduction}
A rewrite $R\to R'$ with the same ordered boundary is an MIS gadget
if and only if, for some constant $c$,
\[
\rvalpha(R')=\rvalpha(R)+c,
\]
where $-\infty+c=-\infty$.
In that case, applying the rewrite to a graph $G$ gives
\[
\mis(G[R\to R'])=\mis(G)+c.
\]
\end{theorem}

\begin{proof}
Let $G$ be any host graph containing $R$. Exterior vertices are adjacent
only to pins, so an independent set of the exterior constrains $R$ in
one way only: the pins adjacent to it cannot be selected. Let
$\mathbf a\in\{0,1\}^k$ mark the pins that may still be selected
($a_i=1$), and let
\begin{equation}
 F_R(\mathbf a)=\max_{\mathbf s\preceq\mathbf a}\valpha(R)_{\mathbf s}
 \label{eq:completion}
\end{equation}
be the size of the largest independent set of $R$ that selects no pin
$b_i$ with $a_i=0$.

Fix an independent set \(J\) of the exterior, and let \(a_i(J)=1\) if and only if \(b_i\) has no neighbor in \(J\). Since the exterior connects to \(R\) only through its pins, the largest independent set with exterior restriction \(J\) has size \(|J|+F_R(\mathbf a(J))\). Hence
\[
\mis(G)=\max_J\bigl(|J|+F_R(\mathbf a(J))\bigr),
\]
where \(J\) ranges over all independent sets of the exterior.

If $\mathbf t$ is dropped, some
$\mathbf s\prec\mathbf t$ has $\valpha(R)_{\mathbf s}\ge\valpha(R)_{\mathbf
t}$. Since $\mathbf s$ selects a subset of the pins selected by $\mathbf t$,
every exterior compatible with
$\mathbf t$ is also compatible with $\mathbf s$, so $\mathbf t$ never
beats $\mathbf s$ in any host. Repeating the argument if $\mathbf s$ is
itself dropped,
\begin{equation}
 F_R(\mathbf a)=\max_{\mathbf s\preceq\mathbf a}\rvalpha(R)_{\mathbf s},
 \label{eq:completion-reduced}
\end{equation}
so the reduced tensor determines $F_R$.

($\Leftarrow$) If $\rvalpha(R')=\rvalpha(R)+c$, then
$F_{R'}(\mathbf a)=F_R(\mathbf a)+c$ for every $\mathbf a$, so
$|J|+F_{R'}(\mathbf a(J))=|J|+F_R(\mathbf a(J))+c$ for every exterior
independent set $J$. Maximizing over $J$ gives
$\mis(G[R\to R'])=\mis(G)+c$. The exterior $J$ of an MIS of
$G[R\to R']$ maximizes $|J|+F_{R'}(\mathbf a(J))$, hence also
$|J|+F_R(\mathbf a(J))$.
Completing $J$ optimally inside $R$ gives an MIS of $G$ with the same
exterior.

($\Rightarrow$) Conversely, $F_R$ determines the reduced tensor. Since
$F_R(\mathbf t)=\max\{\valpha(R)_{\mathbf t},\max_{\mathbf s\prec\mathbf
t}F_R(\mathbf s)\}$, Eq.~\eqref{eq:reduced} becomes
\begin{equation}
 \rvalpha(R)_{\mathbf t}=
 \begin{cases}
  F_R(\mathbf t), & F_R(\mathbf t)>\max_{\mathbf s\prec\mathbf t}F_R(\mathbf s),\\
  -\infty, & \text{otherwise}.
 \end{cases}
 \label{eq:reduced-from-F}
\end{equation}
Thus it suffices to show that $F_{R'}-F_R$ is constant. Assume that
$R\to R'$ is an MIS gadget. Every $\mathbf a$ is reached from
$\mathbf 0$ by allowing one pin at a time, so it is enough to show that
allowing one more pin $b_i$ raises $F_R$ and $F_{R'}$ equally. For
$Q\in\{R,R'\}$, write this gain as
\[
 d_Q=F_Q(\mathbf a)-F_Q(\mathbf a-\mathbf e_i)\in\{0,1\},
\]
where $\mathbf e_i$ is the $i$th unit vector, so we must show
$d_R=d_{R'}$. The gain is 0 or 1 because
allowing a pin never hurts and $b_i$ is a single vertex; $d_Q=1$
exactly when $Q$ needs $b_i$ to reach its best size.

Attach two new leaves to each pin $b_j$ with $a_j=0$. An MIS takes the
leaves instead of $b_j$, since swapping a selected $b_j$ for its two
leaves would enlarge the set. Then attach a new vertex $x$ to $b_i$
only. Counted on $Q$ and $x$, the best size is
$1+F_Q(\mathbf a-\mathbf e_i)$ with $x$ and $F_Q(\mathbf a)$ without
$x$; the leaves add the same amount to both. Hence every MIS contains
$x$ if $d_Q=0$, and some MIS omits $x$ if $d_Q=1$. If
$(d_R,d_{R'})=(0,1)$, some MIS of the rewritten graph omits $x$ and thus
agrees on the exterior with no MIS of the original graph, contradicting
the MIS-gadget property. If $(d_R,d_{R'})=(1,0)$, replace $x$ by a path
$b_i\,x\,y$ with $y$ adjacent only to $x$. An MIS takes exactly one of
$x$ and $y$, with best sizes $1+F_Q(\mathbf a-\mathbf e_i)$ and
$1+F_Q(\mathbf a)$, so some MIS takes $x$ exactly when $d_Q=0$: some
rewritten MIS takes $x$, but no original MIS does. Hence $d_R=d_{R'}$,
so $F_{R'}-F_R$ is a constant $c$, and Eq.~\eqref{eq:reduced-from-F}
gives $\rvalpha(R')=\rvalpha(R)+c$.
\end{proof}

\subsection{Related Work}
\label{sec:related}

\paragraph*{Rydberg-array experiments}
Experimental demonstrations have solved both native instances and
tailored layouts for specific source graphs. Pichler \emph{et
al.}~\cite{pichler2018quantum} proposed MIS on unit-disk graphs as the
native problem, and Ebadi \emph{et al.}~\cite{ebadi2022quantum} solved
native King's-lattice instances with up to 289 atoms. Classical solvers
handle such quasi-planar instances with thousands of vertices within
minutes~\cite{andrist2023hardness}, motivating encodings of harder
nonnative source graphs. Kim \emph{et al.}~\cite{kim2022rydberg} joined
distant atoms by chains of blockaded atoms, called quantum wires, to
realize nonplanar and high-degree graphs. Byun \emph{et
al.}~\cite{byun2022finding} used such wires to embed Platonic graphs in
the plane, Jeong \emph{et al.}~\cite{jeong2023quantum} encoded 3-SAT
instances with dimers, trimers, and wires, and de Oliveira \emph{et
al.}~\cite{de2025demonstration} realized weighted encodings with local
light shifts. While these experiments establish the building blocks for
nonnative encodings on individual graphs, scaling to arbitrary instances
calls for a general automated compiler.

\paragraph*{Encoding arbitrary graphs}
General gadget-based compilers systematically map arbitrary graphs onto
Rydberg arrays. Nguyen \emph{et al.}~\cite{nguyen2023quantum}
introduced the crossing lattice, embedding every graph into a weighted
King's subgraph with quadratic overhead. Liu \emph{et
al.}~\cite{liu2024unweighted} removed vertex weights on the King's
lattice using computer-searched gadgets, and Pan \emph{et
al.}~\cite{pan2025encoding} brought weighted compilation to the
triangular lattice with weights from 1 to 4. De Oliveira \emph{et
al.}~\cite{deoliveira2026sparse} encode sparse graphs with weighted
wires, and Ji \emph{et al.}~\cite{ji2026encoding} map circuit
satisfiability to weighted King's gadgets with a placement and routing
step adapted from VLSI design. Building on these advances, this paper
gives the first unweighted encoding of arbitrary graphs on the
triangular lattice. Bringmann \emph{et al.}~\cite{bringmann2017maximum}
proved that independent set on induced subgraphs of the triangular grid
is NP-complete by a reduction from planar graphs of maximum degree
three; because their source graphs are planar, their reduction needs no
crossing gadget.

\paragraph*{Other encoding approaches}
Schuetz \emph{et al.}~\cite{schuetz2025quantum} combine exact graph
reductions with approximate bottom-up and exact top-down embeddings.
Cazals \emph{et al.}~\cite{cazals2025quantum} decompose graphs into
embeddable subgraphs and merge their solutions into heuristic
independent sets. Parity encodings~\cite{lanthaler2023rydberg} realize
weighted optimization problems in the blockade regime. Anchor-atom
encodings use weak interactions to induce effective local fields under
global driving~\cite{lanthaler2026global}. Local light-shift
encodings~\cite{goswami2024solving} operate outside the blockade regime.
Stastny \emph{et al.}~\cite{stastny2023functional} showed that planar
blockade structures can realize any set of states defined by local
constraints, using minimal logic primitives. Our construction combines
exact decoding with uniform detuning on the triangular lattice.

\paragraph*{Graph drawing and gadget search}
Our placement problems follow VLSI graph layout, and our gadget search
follows exact circuit search. Valiant~\cite{valiant1981universality}
showed that some graphs need quadratic layout area and that planar
layouts may need crossovers. The planarization
approach~\cite{gutwenger2005,tamassia1987} draws a graph with few
crossings and then makes the drawing orthogonal and compact. Our
planarization placement adapts this approach to gadgets of fixed size.
Exact SAT-based search finds minimal Boolean
circuits~\cite{kojevnikov2009finding,soeken2018practical}. Our gadget
search uses SAT in the same way to find small sets of atoms with a given
reduced $\alpha$-tensor.

\section{Problem Formulation}
\label{sec:formulation}

\begin{problem}[Compilation]
\label{prob:compile}
The input is a graph \(G=(V,E)\). The output is a set \(S\) of
triangular-lattice sites whose induced subgraph \(H_S\) satisfies
\(\mis(H_S)=\mis(G)+c\) for a known constant \(c\), and every MIS of
\(H_S\) determines an MIS of \(G\). The
goal is to minimize \(|S|\).
\end{problem}

\begin{figure*}[t]
\centering
\includegraphics{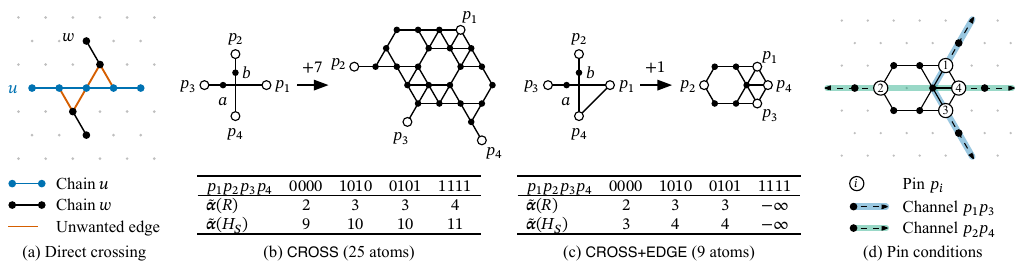}
\caption{\label{fig:gadgets}%
Crossing gadgets. (a) Direct crossings create unwanted edges
between chains. (b,~c) Logical patterns (left) and gadget atoms (right), with
pins as open circles. The tables give the reduced tensors of the
pattern, $\rvalpha(R)$, and of the gadget, $\rvalpha(H_S)$;
unlisted configurations are dropped by reduction, and $1111$ is
infeasible for
\crossedge{}. The indicated MIS offset holds for every exterior.
(d) Pin conditions (G1) and (G2) (Sec.~\ref{sec:search}) on \crossedge{}: the first site (dot)
and exit direction (dashed) of each pin, highlighted by channel; the two
channels cross.}
\end{figure*}

We adopt chain-and-gadget layouts, following the crossing-lattice
strategy of Nguyen \emph{et al.}~\cite{nguyen2023quantum}. Since an atom
has only six lattice neighbors, each source vertex is represented by a
\emph{chain}, or wire, a path of atoms that carries the selection of
that vertex to several locations. The chains of every two vertices cross
once, which gives a location for every possible source edge. A direct
crossing creates unwanted edges between the chains
[Fig.~\ref{fig:gadgets}(a)], so each crossing uses a gadget: \cross{}
lets the two chains pass independently, and \crossedge{} also realizes
the edge between them.

Setting all vertex weights to one invalidates the weighted triangular
gadgets of Pan \emph{et al.}~\cite{pan2025encoding}, and the unweighted
King's-lattice crossing gadget of Liu \emph{et
al.}~\cite{liu2024unweighted} does not embed on the triangular lattice.
Unweighted triangular-lattice compilation therefore requires new
crossing gadgets.

Figure~\ref{fig:flow} summarizes the compilation flow. Gadget search
(Sec.~\ref{sec:search}) finds the crossing gadgets of
Sec.~\ref{sec:composition}, and placement uses either the general
crossing-lattice construction (Sec.~\ref{sec:grid}) or
planarization placement (Sec.~\ref{sec:planar}), a specialized
method for graphs with few crossings.

\section{Gadgets}
\label{sec:composition}

A \emph{chain-and-gadget layout} is built from chains of COPY expansions
and two crossing gadgets. The
crossing gadgets satisfy Theorem~\ref{thm:reduction}, preserving the
reduced $\alpha$-tensor of their boundary up to a constant. The copy gadget of Liu \emph{et al.}~\cite{liu2024unweighted}
distributes the selection of a source vertex to several locations. Where
the copy gadgets of two vertices cross, the \cross{} gadget lets them pass
without interacting, and the \crossedge{} gadget also realizes the source
edge between them. 
Section~\ref{sec:search} describes how we find the two crossing gadgets.

\subsection{Copy Gadget}
\label{sec:copy}

\begin{definition}[COPY gadget~\cite{liu2024unweighted}]
\label{def:copy}
For an integer $k\ge0$, $\mathrm{COPY}_k$ is the path on vertices
$p_1,\ldots,p_{2k+1}$, with the odd-indexed vertices (white) as its
boundary. The rewrite
\begin{equation}
\label{eq:copy}
\vcenter{\hbox{\includegraphics{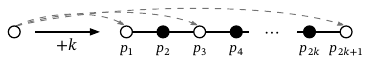}}}
\end{equation}
replaces a single source vertex by this path; the dashed arrows
give the one-to-many boundary correspondence. Each edge incident to the
source vertex is reattached to any one of the boundary vertices, called
its \emph{copies}.
\end{definition}

This COPY rewrite adds $k$ to the MIS size.
For each source vertex $v$, write $A_v$ for the copies of its copy
gadget and $B_v$ for its even-indexed \emph{middle vertices}.
Then $|A_v|=|B_v|+1$.
A path with $2k+1$ vertices has no independent set larger than $k+1$. The only one of size $k+1$ is
$A_v$, the odd positions. Selecting all copies therefore contributes one more
atom than any other choice; this extra atom represents selecting $v$ in $G$. Using the copy gadget, we can propagate the selection of a source vertex to multiple locations in the layout.

\begin{figure}[t]
\centering
\includegraphics{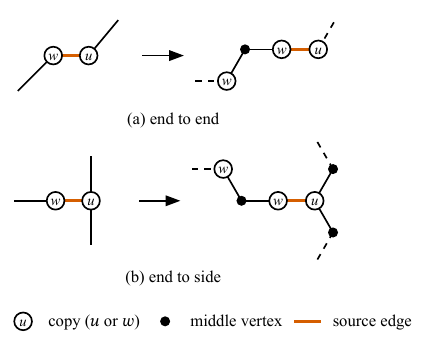}
\caption{\label{fig:joint}%
Direct edges. (a) End-to-end. (b) End-to-side. The only edge between
the lines (orange) joins two copies.}
\end{figure}

An edge between two vertices is usually realized where their wires cross
(Sec.~\ref{sec:cross}). If a wire ends at its neighbor, the wires need not cross. By Definition~\ref{def:copy}, the source
edge $uv$ may be attached to any copy of $u$ and any copy of $v$, so we simply place
the two chains with one copy of each adjacent (Fig.~\ref{fig:joint}). This has two forms. In the \emph{end-to-end} form, both wires end at the
adjacent copies. In the \emph{end-to-side} form, one wire ends next to a copy on a
wire that passes through.

\subsection{Crossing Gadgets}
\label{sec:cross}

The copy gadgets of different vertices must often cross. A direct
crossing places atoms of one chain next to atoms of the other and creates
unwanted edges [Fig.~\ref{fig:gadgets}(a)]. The two \emph{crossing
gadgets}, \cross{} and \crossedge{}, replace such a crossing.

\begin{definition}[\cross{} gadget]
\label{def:cross}
The \cross{} pattern is the graph of disjoint paths $p_1ap_3$ and
$p_2bp_4$ with pins $p_1,\ldots,p_4$. Its two \emph{channels} are the
pin pairs $\{p_1,p_3\}$ and $\{p_2,p_4\}$; each channel carries the chain
of one source vertex through the gadget. The \cross{} gadget is a set of 25
lattice sites with the same pins whose induced graph replaces the pattern
with an MIS offset of 7 [Fig.~\ref{fig:gadgets}(b)].
\end{definition}

By Theorem~\ref{thm:reduction}, this replacement preserves MIS solutions in
every host graph.

When the two crossing vertices are adjacent in $G$, the crossing must also
forbid selecting both.

\begin{definition}[\crossedge{} gadget]
\label{def:crossedge}
The \crossedge{} pattern is the \cross{} pattern with the source edge
$p_1p_4$ added. The \crossedge{} gadget is a set of 9 lattice sites with the
same pins whose induced graph replaces the pattern with an MIS offset of 1
[Fig.~\ref{fig:gadgets}(c)].
\end{definition}

Count the middle vertices $B_v$ in the logical COPY networks before
replacing crossing patterns by physical gadgets. The layout's offset
$c(G)$ is then the sum of $|B_v|$ over source vertices $v$, plus
$7$ per \cross{} and $1$ per \crossedge{}. An MIS of the layout then
has size $\mis(G)+c(G)$.

\section{Gadget Search}
\label{sec:search}

\subsection{Search Problem}

We obtained the \cross{} gadget through an iterative AI-agent research
loop combining SAT-based search and local graph rewrites, and
independently verified every accepted gadget by recomputing its reduced
$\alpha$-tensor and testing its lattice geometry.

We formulate gadget design as a search problem. Given a logical pattern
$R$ and an offset $c$, the search seeks a small set of occupied sites in a
finite triangular-lattice region $W$ that can replace $R$.

The input patterns are the \cross{} and \crossedge{} patterns of
Sec.~\ref{sec:composition}, each with four pins.

The output is a site set $S\subseteq W$ and an ordered set of pins
$P=(p_1,\dots,p_4)$ in $S$. The induced graph $H_S$ must be connected and, with boundary $P$, satisfy
\(
\rvalpha(H_S)=\rvalpha(R)+c.
\)
By Theorem~\ref{thm:reduction}, this tensor constraint ensures that
replacing $R$ by $H_S$ preserves MIS solutions in every host graph and
shifts the optimum by $c$.

The geometric constraints ensure that chains can attach at the pins
without unwanted edges. By the definition of a rewrite, the rest of the graph connects to
a gadget only through its pins. To model the first step
of each connection, we choose an \emph{exit direction} $d(p)$ for each
pin $p$, one of the six triangular-lattice unit vectors. It places the
pin's external neighbor at the \emph{first site} $p+d(p)$.
The pins and exit directions must satisfy two \emph{pin conditions}
[Fig.~\ref{fig:gadgets}(d)]:
\begin{itemize}
\item[(G1)] the first sites lie outside $S$ and are pairwise
  nonadjacent, and each is adjacent to exactly one site of $S$, its own
  pin;
\item[(G2)] the two channels cross: going around the gadget, the first
  sites of $p_1,p_3$ alternate with those of $p_2,p_4$.
\end{itemize}

The search reserves a straight lattice ray from each first site in its
exit direction. Different rays may not intersect or touch, and a ray may
not touch another pin. Sites on or adjacent to a ray are forbidden to
gadget atoms, except its own pin. These restrictions are stricter than (G1).

\subsection{SAT Encoding}
\label{sec:encoding}

The enumeration search space grows exponentially with the number of sites.
We use SAT to search for connected site sets satisfying the tensor constraint.

For fixed pins $P$, exit directions $d$, atom count $N$ and offset $c$,
let $x_v=1$ mean that an atom is placed at allowed site $v$, and let
$\mathbf s\in\{0,1\}^4$ denote a boundary configuration of the four pins. The
formula $\Phi$ enforces four constraints:
\begin{enumerate}
\item \emph{Atom placement.} Place atoms at every pin and at exactly $N$
sites in total:
\[
x_p=1\quad(p\in P),\qquad \sum_v x_v=N.
\]

\item \emph{Connectivity.} The induced graph $H_S$ must be connected.

\item \emph{MIS lower bounds.} For each boundary configuration $\mathbf s$ with finite
$\rvalpha(R)_{\mathbf s}$, require
\[
\valpha(H_S)_{\mathbf s}\ge\rvalpha(R)_{\mathbf s}+c.
\]
\item \emph{MIS upper bounds.} For every boundary configuration $\mathbf s$,
require
\[
\valpha(H_S)_{\mathbf s}\le
\max_{\mathbf t\preceq\mathbf s}\rvalpha(R)_{\mathbf t}+c.
\]
\end{enumerate}

Section~\ref{S-sup:search} of the Supplementary Material describes the
encoding and how each gadget found is verified.

\subsection{SAT Search and Gadget Optimization}
\label{sec:optimization}

We use the formula $\Phi$ from Sec.~\ref{sec:encoding} in a two-stage
search: first find a valid gadget, then shorten its pins.

\paragraph*{Stage I: SAT search}
The outer loop increases the atom count $N$. For each $N$, the inner
loop enumerates pin positions $P$ and exit directions $d$ up to
symmetries of $W$, and solves the corresponding $\Phi$.

\paragraph*{Stage II: Optimization}
Shortening a pin moves it, so the result can lie outside the pin
choices of Stage~I. Starting from the Stage~I gadget, we try two rules:
\begin{enumerate}
\item \emph{Shorten one pin tail.} If a pin $p$ has only one neighbor
$u$, and $u$ has only one other neighbor $v$, delete $p$ and $u$.
Make $v$ the new pin, keeping the label of $p$.
\item \emph{Shorten both ends of a channel.} If both pins of a channel
($\{p_1,p_3\}$ or $\{p_2,p_4\}$) have only one neighbor each, delete the two pins. Make each pin's
neighbor the new pin, keeping its label.
\end{enumerate}

We repeat the rules until neither applies.

\section{Complexity Analysis}
\label{sec:lower}

To establish an optimality benchmark for chain-and-gadget layouts, this
section proves an unconditional $\Omega(n^2)$ worst-case lower bound on
the atom count for $n$-vertex graphs. For an $n$-vertex graph $G$,
let the \emph{separator size} $s(G)$ be the
size of a smallest vertex set $U$ such that every connected component
of $G-U$ has at most $2n/3$ vertices. The following lemma bounds the
size of every layout of $G$ by $s(G)$.

\begin{figure}[t]
\centering
\includegraphics{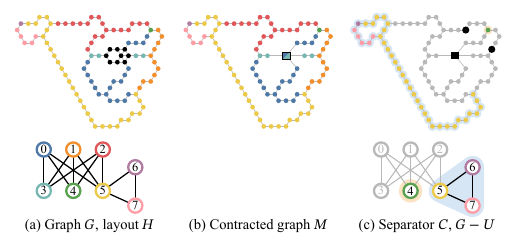}
\caption{\label{fig:skeleton}%
An example of the proof of Lemma~\ref{lem:separator}. (a) Source graph $G$
with $n=8$ and its layout $H$ with 101 atoms. Chain atoms are colored
by source vertex, and the 9 atoms of the \crossedge{} gadget are black.
(b) Contracting the gadget gives $M$: its atoms become one vertex
(split square) shared by vertices 0 and 3, whose chains cross in it.
(c) The separator $C$ (black) consists of the gadget vertex and one atom
each of vertices 1 and 2, so $U=\{0,1,2,3\}$ (gray) and
$|U|=4\le2|C|$. Removing $C$ splits $M$ into two components, of weights
3 and 1 with the weights of $U$ placed on $C$. Each remaining chain
(colored) lies in one of them: those of 5, 6 and 7 in one (blue) and that
of 4 in the other (orange). Hence $G-U$ splits in the same way, into
$\{5,6,7\}$ and $\{4\}$.}
\end{figure}

\begin{lemma}
\label{lem:separator}
Every chain-and-gadget layout of a graph $G$ has at least $s(G)^2/32$ atoms.
\end{lemma}

\begin{proof}
Figure~\ref{fig:skeleton} illustrates the proof on a small example.
Let $H$ be a layout of $G$ with $N$ atoms
[Fig.~\ref{fig:skeleton}(a)]. Each atom of $H$ belongs either to a
crossing gadget or to the chains of exactly one source vertex. The atoms
of each crossing gadget induce a connected subgraph
[Fig.~\ref{fig:gadgets}(b,~c)], and contracting each gadget to a single
vertex gives a graph $M$
[Fig.~\ref{fig:skeleton}(b)]. The graph $H$ is an induced subgraph of the
triangular lattice and hence planar, so $M$ is planar with at most $N$
vertices. For each source vertex $v$, let $M_v$ be the subgraph of $M$
induced by the atoms of the chains of $v$ and the gadgets on them. It is
connected, because the chains of $v$ are joined through these gadgets.
Every vertex of $M$ lies in at most two of these subgraphs: an atom
lies only in $M_v$ for the source vertex $v$ whose chains contain it, and
a gadget lies in $M_u$ and $M_v$ for the two source vertices $u,v$ whose
chains cross in it. Put weight one for each source vertex
on a vertex of $M_v$. The weighted planar separator theorem~\cite{lipton1979separator}
gives a set $C$ of at most $2\sqrt2\sqrt N$ vertices of $M$ such that
every component of $M-C$ has weight at most $2n/3$.

Let $U$ contain every source vertex $v$ whose $M_v$ meets $C$
[Fig.~\ref{fig:skeleton}(c)]. Each vertex of $C$ adds at most two source
vertices to $U$, so $|U|\le2|C|$. For $v\notin U$, $M_v$ lies in one component of $M-C$. The layout
realizes each source edge $uv$ either by a \crossedge{} shared by $M_u$
and $M_v$ or by a direct edge of $H$ between them (Fig.~\ref{fig:joint}). Hence, if $uv\in E$ with
$u,v\notin U$, then $M_u$ and $M_v$ lie in the same component. Thus every
component of $G-U$ has at most $2n/3$ vertices [Fig.~\ref{fig:skeleton}(c)], and
$s(G)\le|U|\le2|C|\le4\sqrt2\sqrt N$, that is, $N\ge s(G)^2/32$.
\end{proof}

\begin{theorem}[Worst-case size]
\label{thm:lower}
Compiling $n$-vertex graphs into chain-and-gadget layouts requires
$\Omega(n^2)$ atoms in the worst case.
\end{theorem}

\begin{proof}
By Lemma~\ref{lem:separator}, it suffices to exhibit $n$-vertex graphs
with $s(G)=\Omega(n)$. As $n\to\infty$ over even $n$, a uniformly random
3-regular graph $G$ has, with probability tending to one, edge expansion
at least a constant $h>0$: every set $X$ of at most $n/2$ vertices has
at least $h|X|$ edges leaving
it~\cite{friedman2008proof,fiedler1973algebraic}. Let $U$ be any vertex
set such that every component of $G-U$ has at most $2n/3$ vertices, and
suppose $|U|\le n/10$. If some component of $G-U$ has at least $n/4$
vertices, let $S$ be that component. Otherwise, add components to $S$
one at a time until $|S|\ge n/4$, which ends with $|S|<n/2$. In both
cases $n/4\le|S|\le2n/3$, so the smaller of $S$ and $V\setminus S$ has
at least $n/4$ vertices and at least $hn/4$ edges leave $S$. Every edge
leaving $S$ ends in $U$, so at most $3|U|$ edges leave $S$, which gives
$|U|\ge hn/12$. Hence $s(G)\ge\min(1/10,h/12)\,n$, and by
Lemma~\ref{lem:separator} every layout of $G$ has $\Omega(n^2)$ atoms.
\end{proof}

\section{Crossing-Lattice Placement}
\label{sec:grid}

\begin{figure*}[t]
\centering
\includegraphics[width=\textwidth]{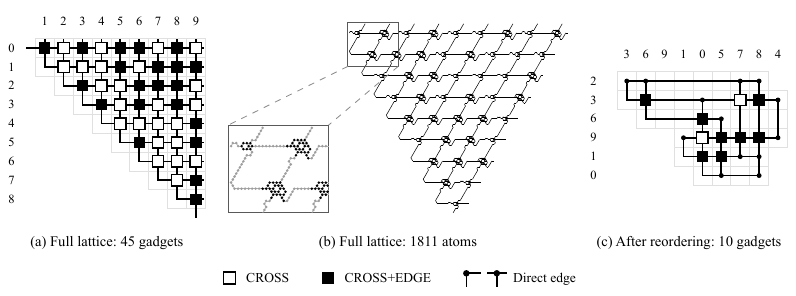}
\caption{\label{fig:grid}%
From the full crossing lattice to the crossing-lattice placement for the
example graph of Fig.~\ref{fig:flow}. (a) Full lattice: each interior wire
runs down its column, turns once, and runs along its row; the first and
last wires are straight. Every pair crosses once, in
a \crossedge{} (filled) or \cross{} (open). (b) The full lattice has
1811 atoms. The inset magnifies the boxed region, which covers the four
cells in rows 0--1 and columns 1--2 of (a): crossing gadgets (black)
and the chain atoms joining them (gray). (c) After the reordering and
trimming of Sec.~\ref{sec:grid-steps}, each line spans only its first to last neighbor, and only 10
gadgets remain. Line ends meet in end-to-end or end-to-side direct edges (dots).
The placement has 510 atoms (Fig.~\ref{fig:flow}, right).}
\end{figure*}

This is the simplest placement method. It follows the crossing lattice
of~\cite{nguyen2023quantum,pan2025encoding} and works for any graph. Draw each vertex as a COPY gadget, i.e., an atom wire. As shown in Fig.~\ref{fig:grid}(a), wire $v$ occupies column $v$ above the diagonal and row $v$
to its right, with a bend joining the two parts when both are present. Any two wires $i<j$
then cross exactly once, in the grid cell $(i,j)$. The graph is encoded by
what we put in that cell:
\begin{itemize}
\item if $ij$ is an edge, the cell holds a \crossedge{} gadget: the two
wires cross, and one edge joins them;
\item if $ij$ is not an edge, the cell holds a \cross{} gadget: the two
wires cross without touching.
\end{itemize}
Every pair gets a gadget, even when a wire has no neighbors left to meet.
Section~\ref{sec:grid-steps} removes the unneeded gadgets [Fig.~\ref{fig:grid}(c)].

\subsection{Construction}

Number the vertices $0,\dots,n-1$. Each vertex $v$ becomes an L-shaped
COPY chain (Definition~\ref{def:copy}), called wire $v$: it goes down column
$v$ through rows $0,\dots,v-1$, bends in cell $(v,v)$, and goes along row $v$
through columns $v+1,\dots,n-1$. Wires $0$ and $n-1$ are straight. In this
full lattice, every two wires $i<j$ cross exactly once, in cell $(i,j)$, and we
place a gadget there according to the adjacency of $i$ and $j$: a \crossedge{}
if $ij\in E$ and a \cross{} otherwise (Algorithm~\ref{alg:lattice}). The
copies of each vertex must appear exactly at the gadget pins, so the
length of each wire segment must have the right parity. Where a straight segment
would have the wrong parity, the chain takes a one-site detour, which adds a
small bend [Fig.~\ref{fig:grid}(b), inset].

\begin{algorithm}[t]
\caption{Crossing-lattice placement}
\label{alg:lattice}
\begin{algorithmic}[1]
\Require graph $G=(V,E)$ with $V=\{0,\dots,n-1\}$ and $n\ge2$
\Ensure layout $H$
\State $H\gets\emptyset$
\For{each vertex $v$}
  \State lay an L-shaped wire for $v$: down column $v$, bending at cell $(v,v)$, then along row $v$
\EndFor
\For{each pair $0\le i<j\le n-1$}
  \If{$ij\in E$}
    \State put a \crossedge{} gadget in cell $(i,j)$, where wires $i$ and $j$ cross
  \Else
    \State put a \cross{} gadget in cell $(i,j)$, where wires $i$ and $j$ cross
  \EndIf
\EndFor
\State \Return $H$
\end{algorithmic}
\end{algorithm}

\subsection{Complexity Analysis}
Every COPY wire and crossing gadget preserves MIS solutions by
Definition~\ref{def:copy} and Theorem~\ref{thm:reduction}, so we focus
on bounding the atom count.
\begin{theorem}[Upper bound of the crossing-lattice placement]
\label{thm:grid}
For every graph $G$ on $n\ge2$ vertices, Algorithm~\ref{alg:lattice}
constructs an induced graph $H$ on the triangular lattice
with at most $21.5\,n^2$ atoms. 
\end{theorem}

\begin{proof}
Let $m=|E|$ and $X=\binom n2-m$ count adjacent and nonadjacent pairs.
A cell holds its gadget and the wire atoms inside it: 41 atoms for a
\cross{} and 31 for a \crossedge{}. Each of the $n-2$ bends has 11 atoms.
One more atom joins each two consecutive pieces of a wire, which gives
$n-1$ such atoms on each bent wire and $n-2$ on each straight one, or
$(n-2)(n+1)$ in total. Summing and using $X+m=\binom n2$,
\begin{equation}
\begin{aligned}
 N_{\mathrm{full}}&=43X+33m+11n-24\\
 &=\tfrac{43}{2}n^2-10m-\tfrac{21}{2}n-24<21.5\,n^2,
\end{aligned}
 \label{eq:grid}
\end{equation}
for $n\ge2$.
\end{proof}

\subsection{Reducing the Overhead}
\label{sec:grid-steps}

The full lattice places a gadget in every cell, even beyond a wire's
active span. Shortening each wire to run only from its first to its last
neighbor eliminates crossings between non-overlapping wires and leaves
their cells empty. How many wires overlap depends on the vertex order.
We use a greedy vertex order and refine it by moving individual vertices
to new positions in the order. For example, the graph of
Fig.~\ref{fig:grid} needs only 10 gadgets after reordering and shortening,
and its layout shrinks from 1811 to 510 atoms.

If a wire ends at its neighbor, the two wires need not cross, and we join them by a
direct edge (Sec.~\ref{sec:copy}) instead of a crossing gadget. Finally, we remove
redundant atom pairs while preserving the independent-set choices allowed
at the ends of each connecting path.

\section{Planarization Placement}
\label{sec:planar}

Every crossing of two chains costs one crossing gadget, and the
crossing lattice crosses two wires wherever their spans overlap.
Planarization placement instead derives the layout from a drawing of $G$
with few crossings. For the graph of Fig.~\ref{fig:flow}, planarization
needs only 2 crossings, compared with 10 for the crossing lattice.

A drawing may require a COPY gadget that can branch. 

\begin{definition}[\texorpdfstring{\branchcopy{}}{BRANCH-COPY} gadget]
\label{def:tree}
A \branchcopy{} gadget is a COPY gadget that may branch at its
copies (white). The rewrite
\begin{equation}
\label{eq:tree}
\vcenter{\hbox{\includegraphics{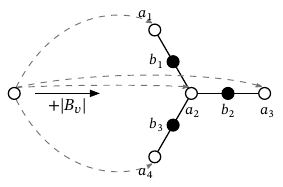}}}
\end{equation}
replaces a single source vertex by this gadget. Each edge incident to
the source vertex is reattached to any one of the boundary vertices, its copies.
\end{definition}

As for COPY, the copies $A_v$ are the only largest independent set of
the gadget, so the rewrite adds $|B_v|$ to the MIS size, where $B_v$
is the set of middle vertices (black) between the copies.

\subsection{Construction}
\begin{figure*}[t]
\centering
\includegraphics[width=\textwidth]{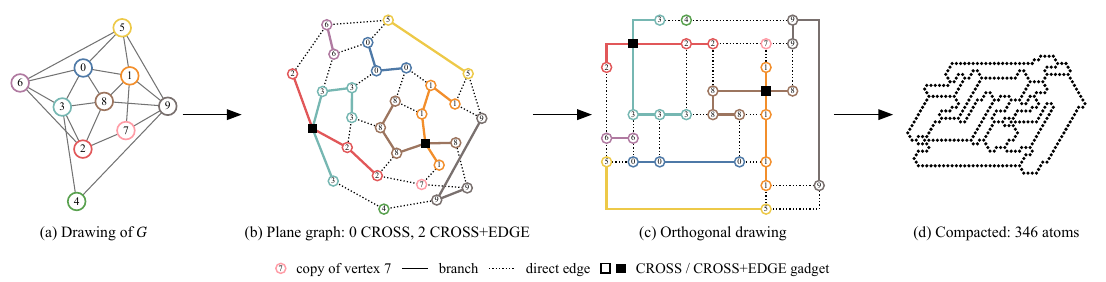}
\caption{\label{fig:stages}%
Planarization placement on the example graph. (a) Straight-line drawing
found by planarize, with two crossings. For legibility it is spread
without changing its plane graph. (b) Plane graph after planarize, spread in the same
way. Each vertex becomes a copy gadget, whose
copies are numbered by vertex and joined by branches (solid), with one
color per vertex as in (a). Each edge keeps one direct edge (dotted)
between copies of its two vertices. Each crossing becomes a \cross{}
(open square), or a \crossedge{} (filled) if the two vertices are
adjacent. (c) After shape: the orthogonal drawing of the same plane
graph, with its angles, bends and coordinates. Mapped to the lattice, it
gives a layout with 2888 atoms. (d) After compact: every chain
shortened by minimum-cost flows, 346 atoms against 510 for the
crossing-lattice placement.}
\end{figure*}

Write $T_u$ for the COPY or \branchcopy{} gadget representing source
vertex $u$. Algorithm~\ref{alg:planar} has three steps (Fig.~\ref{fig:stages}).
\emph{Planarize} builds a plane graph $\mathcal N$, a graph drawn without
crossings, whose nodes are copies and crossing gadgets. \emph{Shape}
redraws $\mathcal N$ on the lattice, keeping the order in which the edges
leave each node, so the result is also free of crossings.
\emph{Compact} shortens the chains.

\begin{algorithm}[t]
\caption{Planarization placement}
\label{alg:planar}
\begin{algorithmic}[1]
\Require graph $G=(V,E)$
\Ensure layout $H$
\State $D\gets$ drawing of $G$ with few crossings
\State $\mathcal N\gets D$ with COPY or \mbox{\branchcopy{}} gadgets at
\Statex \hspace{\algorithmicindent}vertices and crossing gadgets at crossings \Comment{planarize}
\State $H\gets$ orthogonal drawing of $\mathcal N$ on the lattice \Comment{shape}
\State shorten the chains of $H$ \Comment{compact}
\State \Return $H$
\end{algorithmic}
\end{algorithm}

\paragraph*{Planarize}
Minimizing crossings is NP-hard~\cite{garey1983crossing}, so a nonplanar
graph is drawn by simulated annealing~\cite{kirkpatrick1983optimization},
starting from a force-directed layout~\cite{fruchterman1991graph} on an
integer grid. In each sweep, every vertex is tried at 23 candidate
points: the centroid of its neighbors, perturbations of six random
neighbors and of six random vertices, and ten uniform points in the
bounding box. The best candidate is taken if it removes crossings;
otherwise a random candidate that adds $\Delta$ crossings is accepted
with probability $e^{-\Delta/T}$, where $T$ decreases linearly to zero.
The drawing with the fewest crossings is kept. 

The plane graph $\mathcal N$ turns the drawing into chains and gadgets
[Fig.~\ref{fig:stages}(a,~b)]. The chain $T_u$ of each vertex $u$ starts
at $u$ and runs along each of its edges toward the middle. On an edge
$uv$, the chains of $u$ and $v$ meet at the segment between crossings
closest to the middle of $uv$, which becomes a direct edge between a
copy of $u$ and a copy of $v$. Where the chains of two vertices $x$ and
$y$ cross, $\mathcal N$ places a \crossedge{} if $xy\in E$ and a
\cross{} otherwise. Copies with more than three neighbors are split,
leaf copies are removed, and two adjacent copies are merged if the
result has at most three neighbors, none repeated. Hence $T_u$ and
$T_v$ are joined exactly when $uv\in E$, by a direct edge and possibly a
\crossedge{} that imposes the same constraint.

\paragraph*{Shape}
Tamassia's algorithm~\cite{tamassia1987} computes the angles and bends
of an orthogonal drawing that keeps this edge order as a
minimum-cost flow, in which a bend, or a turn at a node of degree two,
costs one. The two drawing axes map to lattice directions $60^\circ$
apart and drawing coordinates to even lattice coordinates, so each bend
becomes a $60^\circ$ or $120^\circ$ turn. Each node becomes a fixed set
of atoms, and each chain leaves it as one of a finite catalog of arms;
we verified by computer that every combination induces exactly the
intended edges. The copies of a chain must sit at its odd positions, so shape also gives
each chain a length of the right parity
(Supplementary Sec.~\ref{S-sec:planar-shape}).

\paragraph*{Compact}
Compact shortens the chains without changing the shape of the drawing,
as in constraint-graph compaction of VLSI
layouts~\cite{lengauer1990,klau1999}. Along each lattice axis, the chain
segments parallel to it may shrink while the rest of the layout moves
rigidly. Minimizing their total length, with atoms kept apart, is the
dual of a minimum-cost flow and is solved exactly. The passes repeat
until no chain gets shorter, and every chain keeps its parity
(Supplementary Sec.~\ref{S-sup:compact}).

\subsection{Complexity Analysis}
\label{sec:benefit}

The crossing-lattice and planarization placements complement each other
across graph families. Two parameters of $G$ govern layout size: the
crossing number $\mathrm{cr}(G)$, the fewest edge crossings in any plane
drawing, and the separator size $s(G)$ of Sec.~\ref{sec:lower}.
Algorithm~\ref{alg:lattice} guarantees at most $21.5\,n^2$ atoms on
every graph (Theorem~\ref{thm:grid}). Planarization replaces each
crossing with a gadget of at least 9 atoms, requiring at least
$9\,\mathrm{cr}(G)$ atoms, while Lemma~\ref{lem:separator} gives the
universal lower bound $s(G)^2/32$.

A planar graph has no crossings and
$s(G)=O(\sqrt n)$~\cite{lipton1979separator}, and a $k\times k$ torus
grid has fewer than $n$ crossings~\cite{harary1973toroidal} and
$s(G)\le2\sqrt n$, so neither bound stands in the way of layouts far
smaller than $n^2$. Table~\ref{tab:structured} shows the largest
savings on these graphs.

On sparse expanders, both placements require at least quadratically many atoms. With probability
tending to one, a random 3-regular graph is an
expander~\cite{friedman2008proof}: every set of at most $n/2$ vertices
has a constant fraction of its size in outgoing edges. Hence
$s(G)=\Theta(n)$ and $\mathrm{cr}(G)=\Theta(n^2)$~\cite{pach1996applications},
every layout needs $\Omega(n^2)$ atoms, and planarization saves at most
a constant factor.

On dense graphs, the crossing lattice is asymptotically smaller. The
crossing lemma~\cite{ajtai1982crossing,leighton1984new},
$\mathrm{cr}(G)\ge m^3/(64n^2)$ for $m\ge4n$, makes planarization larger
than the crossing lattice whenever $m\ge5.4\,n^{4/3}$. On $K_{n/2,n/2}$,
and with probability tending to one on $G(n,p)$ with fixed $p$,
planarization needs $\Omega(n^4)$ atoms against $O(n^2)$.

\section{Experimental Results}
\label{sec:results}

Our experiments evaluate both compilation overhead and quantum annealing
dynamics. First, the crossing lattice compiles every tested graph within
its $O(n^2)$ bound, while planarization yields much smaller layouts on
sparse and low-crossing graphs (Sec.~\ref{sec:scaling}). Second, in
coherent quantum annealing simulations of $K_{3,3}$, longer sweeps
rapidly suppress blockade violations and steadily raise the direct MIS
success probability on the triangular layout without postprocessing,
whereas on the King's layout violations persist and the success
probability fails to grow even with postprocessing (Sec.~\ref{sec:gap}).

We test sparse random 3-regular graphs, dense Erd\H{o}s--R\'enyi
graphs $G(n,0.3)$, and four structured graph families, using
King's-lattice layouts generated by \texttt{pred}~\cite{pan2026problem}
as a baseline.

\subsection{Scalability}
\label{sec:scaling}

\begin{figure*}[t]
\centering
\includegraphics[width=\textwidth]{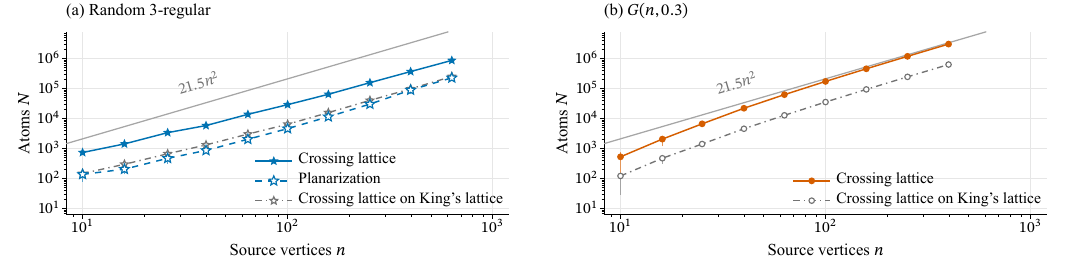}
\caption{\label{fig:scaling}%
Atoms $N$ against source vertices $n$ for (a) random 3-regular graphs
and (b) $G(n,0.3)$: medians of ten graphs, with bars from minimum to
maximum. The crossing lattice omits the optional vertex-order refinement and pair
removal of Sec.~\ref{sec:grid-steps}. Planarization is shown on
3-regular graphs only. The gray dash-dotted series is the
crossing-lattice method on King's lattice after post-optimization, up to
$n=632$ in (a) and $n=398$ in (b). The light gray line is the worst case
$21.5n^2$ of Theorem~\ref{thm:grid}.}
\end{figure*}

The crossing lattice stays within its bound on every graph
(Fig.~\ref{fig:scaling}). On 3-regular graphs, reordering and shortening
the wires remove most gadgets. A least-squares fit to the plotted median
atom counts for $40\le n\le632$ gives
$N_{\mathrm{lattice}}\approx2.03n^2+123n$,
far below the worst case $21.5n^2$. On dense $G(n,0.3)$ graphs, almost
every pair of wires must cross, so the layouts grow as about $20n^2$,
close to the worst case.

Planarization gives smaller layouts on the tested random 3-regular
graphs (Fig.~\ref{fig:scaling}). A least-squares fit to the median atom
counts for $40\le n\le632$ gives
$N_{\mathrm{planar}}\approx0.62n^2-22n$.
Both methods show nearly quadratic atom
growth, consistent with the lower bound of Theorem~\ref{thm:lower}.
On dense $G(n,0.3)$ graphs the ratio reverses and grows with $n$:
the median planarization layout is 1.4 times the size of the crossing
lattice at $n=25$ and 4.4 times at $n=40$, as the crossing lemma
predicts (Sec.~\ref{sec:benefit}).

\begin{table}[t]
\centering
\setlength{\tabcolsep}{3pt}
\caption{\label{tab:structured}The number of crossing gadgets and atoms on structured graphs.}
\begin{tabular}{@{}lrrrrrrr@{}}
\toprule
 & & & \multicolumn{2}{c}{Planarization} & \multicolumn{2}{c}{Crossing lattice} & King's \\
\cmidrule(lr){4-5}\cmidrule(lr){6-7}\cmidrule(l){8-8}
Graph & $n$ & $m$ & Gadgets & Atoms & Gadgets & Atoms & Atoms \\
\midrule
Dodecahedron & 20 & 30 & 0 & \textbf{224} & 8 & 1\,136 & 508 \\
$10\times10$ grid & 100 & 180 & 0 & \textbf{2\,828} & 67 & 10\,397 & 4\,824 \\
\midrule
Petersen & 10 & 15 & 2 & \textbf{218} & 3 & 397 & 219 \\
Heawood & 14 & 21 & 3 & \textbf{347} & 6 & 616 & 344 \\
Tutte--Coxeter & 30 & 45 & 14 & \textbf{1\,434} & 25 & 2\,927 & 1\,212 \\
\midrule
$6\times6$ torus grid & 36 & 72 & 24 & \textbf{2\,480} & 51 & 4\,031 & 2\,156 \\
$10\times10$ torus grid & 100 & 200 & 96 & \textbf{11\,890} & 276 & 25\,776 & 9\,657 \\
\midrule
$Q_5$ & 32 & 80 & 56 & \textbf{5\,112} & 87 & 5\,605 & 2\,129 \\
$Q_7$ & 128 & 448 & 2\,208 & 203\,930 & 1\,883 & \textbf{110\,113} & 28\,823 \\
\midrule
$K_{8,8}$ & 16 & 64 & 144 & 8\,934 & 36 & \textbf{1\,052} & 415 \\
$K_{16,16}$ & 32 & 256 & 3\,136 & 173\,158 & 196 & \textbf{4\,540} & 1\,727 \\
\bottomrule
\end{tabular}
\end{table}

Planarization gains most on graphs that can be drawn with few crossings
(Table~\ref{tab:structured}). Planar graphs need no crossing gadget, and
on the dodecahedron and the $10\times10$ grid planarization gives
layouts 5.1 and 3.7 times smaller than the crossing lattice. On the
Petersen, Heawood and Tutte--Coxeter graphs, the torus grids and $Q_5$,
it creates fewer crossing gadgets and gives the smaller layout. On
$K_{8,8}$ and $K_{16,16}$, the drawings have 144 and 3\,136 crossings,
as many as Zarankiewicz's drawing, which is conjectured to be
optimal~\cite{zarankiewicz1954problem}. The crossing lattice is still 8.5
and 38 times smaller, because the crossing number of $K_{n,n}$ grows as
$n^4$ (Sec.~\ref{sec:benefit}).

\subsection{Simulation}
\label{sec:gap}

We simulate coherent quantum annealing on triangular and King's arrays.

\begin{figure*}[t]
\centering
\includegraphics[width=\textwidth]{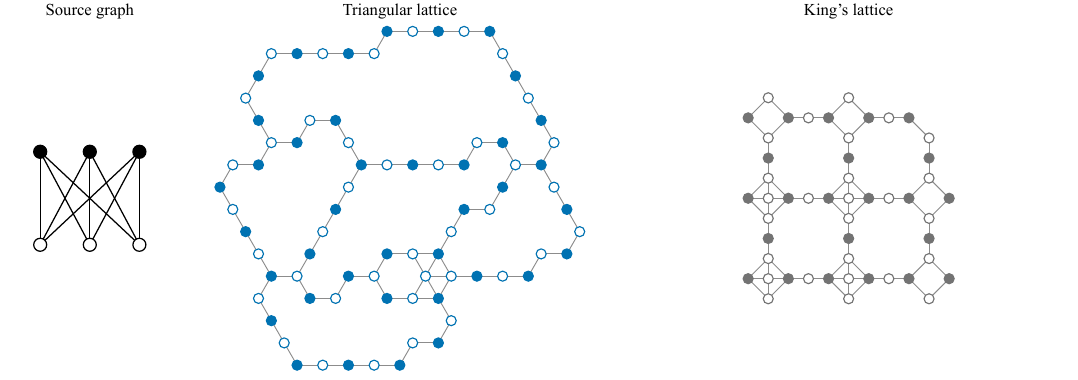}
\caption{\label{fig:k33-layouts}The source graph $K_{3,3}$ and its
79-atom triangular and 50-atom King's layouts, drawn at one length
scale. Filled atoms form a maximum independent set of each layout,
which decodes to the maximum independent set filled in the source
graph.}
\end{figure*}

\paragraph*{Setup and numerical method}
We use the complete bipartite graph $K_{3,3}$, the nonplanar graph with
the fewest edges. It is encoded with 79 atoms on the
triangular lattice by planarization placement, with one crossing gadget,
and with 50 atoms on the King's lattice (Fig.~\ref{fig:k33-layouts};
Supplementary Sec.~\ref{S-sup:large-graphs}). We evolve matrix product states (MPS)
using the time-dependent variational principle (TDVP) with global
subspace expansion~\cite{pan2025encoding}. The maximum bond dimension is
$\chi=64$, and the Hamiltonian of Eq.~\eqref{eq:rydberg} retains every
pair interaction $C_6/r^6$.

\begin{figure*}[t]
\centering
\includegraphics{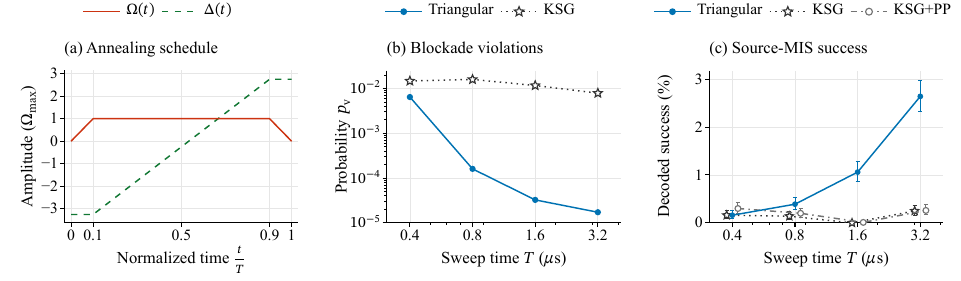}
\caption{\label{fig:large-dyn}%
Quantum annealing of $K_{3,3}$ on the 79-atom triangular and 50-atom
King's arrays. (a)~Annealing schedule shared by both arrays.
(b)~Blockade-violation probability $p_{\mathrm v}$
[Eq.~\eqref{eq:violation-probability}] in the final simulated state,
before classical postprocessing.
(c)~Decoded source-MIS success: direct decoding on the triangular array
and on the King's array (KSG) [Eq.~\eqref{eq:decoded-mis-probability}],
and vertex reduction followed by vertex addition on the King's array
(KSG+PP). Bars are 95\% Wilson intervals for sampling uncertainty over
$10^4$ shots. Markers of
the two King's curves are offset horizontally for legibility; they use
the same quantum samples. A marker at zero means that no shot succeeded.}
\end{figure*}

Both lattices use $C_6=2\pi\times862\,690$~MHz\,$\mu$m$^6$ and
$\Omega_{\max}=2\pi\times4$~MHz~\cite{ebadi2022quantum}. The spacings
are 5.88~$\mu$m on the triangular lattice and 4.60~$\mu$m on the King's
lattice, following the rule of Ref.~\cite{pan2025encoding}. Both follow the schedule in
Fig.~\ref{fig:large-dyn}(a), and we test annealing times $T\in\{0.4,0.8,1.6,3.2\}$~$\mu$s.

\paragraph*{Observables}
Let $H=(V_H,E_H)$ be the layout's blockade graph and $|\psi(T)\rangle$
the final simulated state. The edge-averaged \emph{blockade-violation
probability} is
\begin{equation}
 p_{\mathrm v}(T)=\frac{1}{|E_H|}
 \sum_{(i,j)\in E_H}\langle n_i n_j\rangle_T,
 \label{eq:violation-probability}
\end{equation}
where the expectation is evaluated in $|\psi(T)\rangle$. It measures how
often adjacent atoms are excited together. The \emph{direct decoded MIS
probability} is
\begin{equation}
 P_{\mathrm{MIS}}(T)=
 \sum_{\mathbf z\in\mathcal S_{\mathrm{MIS}}}
 |\langle\mathbf z|\psi(T)\rangle|^2,
 \label{eq:decoded-mis-probability}
\end{equation}
where $\mathcal S_{\mathrm{MIS}}$ contains the occupation strings whose
excited atoms form an independent set of $H$ and decode to an MIS of the
source graph. We estimate $P_{\mathrm{MIS}}$ by the fraction of $10^4$
samples per run that belong to this set.

\paragraph*{Classical postprocessing}
At every $T$, we also apply the vertex-reduction and vertex-addition
procedures of Ref.~\cite{ebadi2022quantum} to the King's samples. Vertex
reduction repeatedly removes the occupied vertex with the most occupied
neighbors, breaking ties randomly, until the physical configuration is
independent. Vertex addition independently augments the repaired
configuration in ten random vertex orders, adding each vertex whenever
independence is preserved. We keep the largest resulting physical
independent set and apply the same decoder. A shot succeeds only if its decoded vertices form a
maximum independent set of the source graph, whose optimum is checked by
exhaustive enumeration.

\paragraph*{Results}
The two layouts show markedly different scaling with annealing time $T$
[Fig.~\ref{fig:large-dyn}(b,~c)]. On the triangular array, $p_{\mathrm v}$
drops sharply as $T$ increases, widening its advantage over the King's
array from 2.3-fold at $T=0.4$~$\mu$s to 454-fold at $T=3.2$~$\mu$s
[Fig.~\ref{fig:large-dyn}(b)]. Correspondingly, the direct decoded MIS
probability $P_{\mathrm{MIS}}$ on the triangular array grows steadily
with $T$ without postprocessing, reaching 2.6\% at $T=3.2$~$\mu$s
[Fig.~\ref{fig:large-dyn}(c)]. On the King's array, by contrast,
$p_{\mathrm v}$ stays on the order of $10^{-2}$ and $P_{\mathrm{MIS}}$
fails to grow with $T$, remaining at or below 0.3\% even after classical
postprocessing. While postprocessing repairs every King's sample into an
independent set of $H$, the repaired configurations rarely decode to an
MIS of $K_{3,3}$.

\paragraph*{Analysis}
The difference in $p_{\mathrm v}$ follows from the energy of a
violation at the end of the sweep. Adding one atom to a maximum
independent set of the layout raises the energy by at least
$7.7\Omega_{\max}$ on the triangular array, compared with $3.0\Omega_{\max}$
on the King's array, whose weakest blockade interaction barely exceeds
the final detuning. Violating configurations therefore remain close in
energy to the MIS on the King's array, so slower sweeps do not remove
them.

On the triangular array, more than 98\% of samples are independent from
$T=0.8$~$\mu$s on, whereas on the King's array at most 53\% are. The
remaining non-optimal samples arise from conflicts between the two
degenerate solutions of $K_{3,3}$: different parts of the array settle
into different solutions, losing an atom where they meet. Longer sweeps
reduce these conflicts and steadily raise $P_{\mathrm{MIS}}$, and
optimizing the global annealing schedule can improve performance further
(Sec.~\ref{sec:discussion}).

\section{Discussion}
\label{sec:discussion}

The quadratic atom-count exponent of our compiler is optimal beyond
chain-and-gadget layouts under the standard assumption
$\mathrm{NP}\not\subseteq\mathrm{coNP}/\mathrm{poly}$.
In particular, no deterministic polynomial-time exact compiler with a
computable MIS offset can guarantee an unweighted triangular-lattice
encoding with $O(n^{2-\varepsilon})$ atoms for arbitrary $n$-vertex source graphs and any
fixed $\varepsilon>0$. Such a compiler would compress a Vertex Cover
instance into an equivalent MIS decision instance of
$O(N\log N+\log n)$ bits, since every output atom has at most six
neighbors. This would contradict known lower bounds on the size of
equivalent instances produced in polynomial
time~\cite{dell2014sparsification}, independently of the exponential time
hypothesis. For chain-and-gadget layouts, Theorem~\ref{thm:lower} gives the
stronger, unconditional $\Omega(n^2)$ bound. Smaller constant factors and
improved performance on structured graphs remain useful directions.

The original optimality argument for the unweighted King's-lattice
reduction~\cite{liu2024unweighted} combined an $N$-atom encoding with
$2^{O(\sqrt N)}$-time tensor contraction to contradict the exponential
time hypothesis. OpenAI's reported refutation~\cite{openai2026eth}
gives a $2^{O(n/\sqrt{\log n})}$-time algorithm for sparse 3-SAT, with a
logarithmic saving in the exponent. The quadratic-exponent barrier still
follows from the compression assumption above, since King's-lattice
graphs also have bounded degree. Logarithmic atom savings outside
chain-and-gadget layouts remain open: a hypothetical $N=O(n^2/\log n)$ encoding
would give $2^{O(n/\sqrt{\log n})}$-time solution by two-dimensional
tensor contraction, matching the reported sparse-SAT exponent scale.
Such logarithmic savings are not ruled out by the compression bound.

To implement the proposed encoding, atoms can be arranged at the compiled
triangular-lattice sites and driven by global Rabi and detuning
schedules~\cite{ebadi2022quantum}. Our unweighted gadgets simplify control
relative to the earlier weighted triangular-lattice encoding~\cite{pan2025encoding}
by allowing the same detuning at every atom. Measurements of
individual COPY and crossing gadgets, followed by complete layouts, can
identify robust operating ranges in lattice spacing and sweep duration.
Experiments on quantum wires highlight the value of testing wire-length
dependence~\cite{kim2022rydberg,bombieri2025quantum}. Optimizing the global
annealing schedules offers a route to improving preparation while
retaining uniform detuning~\cite{finzgar2024designing,perseguers2025hardness}.
Calibrated atom-loss and readout measurements would support a quantitative
comparison of blockade violations and decoded MIS probabilities with our
simulations~\cite{kim2024dataset}.

Beyond this problem, the AI-assisted discovery of \cross{} shows how
agents can combine SAT-based search and local graph rewrites with exact
verification to design reduction gadgets. Adapting the target constraints
and verifier could support similar searches for planarization in
Dominating Set, degree reduction in Feedback Vertex Set, and
solution-count-preserving SAT reductions.

\section{Conclusion}
\label{sec:conclusion}

Suppressing constraint violations natively in hardware is a promising
route toward better performance scaling in Rydberg quantum optimization. Because the triangular lattice separates intended from
unwanted atomic interactions more strongly than the King's lattice, we
developed an unweighted compiler that maps maximum independent set
problems on arbitrary graphs onto triangular Rydberg atom arrays with
uniform detuning and exact solution recovery. Our crossing-lattice
placement guarantees at most $21.5n^2$ atoms for any $n$-vertex graph,
while our planarization placement uses graph structure to produce
smaller layouts on sparse graphs. In coherent quantum annealing
simulations of the nonplanar graph $K_{3,3}$, longer annealing rapidly
suppresses constraint violations on the triangular layout and steadily
raises the direct MIS success probability without postprocessing,
whereas on the King's layout violations persist and the success
probability fails to grow even with the postprocessing of Ebadi
\emph{et al.} These results provide evidence that encoding constraints
cleanly into the triangular lattice can make postprocessing-free quantum
annealing much more efficient.

\section*{Code and Data Availability}
The compiler code is hosted at
\url{https://github.com/wuxian3/triangular-mis}.
We also plan to provide the mapping through the
\href{https://github.com/CodingThrust/problem-reductions}{\texttt{problem-reductions}}
project~\cite{pan2026problem} for use alongside other problem reductions.

\section*{Acknowledgment}
The authors used OpenAI Codex agents in this work. They designed and ran the gadget search
(Sec.~\ref{sec:search}), helped implement the placement algorithms
(Secs.~\ref{sec:grid} and~\ref{sec:planar}) and the 
simulations (Sec.~\ref{sec:gap}). They also
helped draft and edit the manuscript. The authors reviewed
all AI-assisted text, code, and figures, and take full responsibility
for the content of this article.

\bibliographystyle{IEEEtran}
\bibliography{references}

\clearpage
\twocolumn[\centering
  \pdfbookmark{Supplementary Material}{supplementary-material}
  {\LARGE Supplementary Material\par}\vspace{1em}]
\setcounter{section}{0}
\renewcommand{\thesection}{S\arabic{section}}
\renewcommand{\theHsection}{S\arabic{section}}
\setcounter{figure}{0}
\renewcommand{\thefigure}{S\arabic{figure}}
\renewcommand{\theHfigure}{S\arabic{figure}}
\setcounter{table}{0}
\renewcommand{\thetable}{S\arabic{table}}
\renewcommand{\theHtable}{S\arabic{table}}
\setcounter{equation}{0}
\renewcommand{\theequation}{S\arabic{equation}}
\renewcommand{\theHequation}{S\arabic{equation}}
\setcounter{algorithm}{0}
\renewcommand{\thealgorithm}{S\arabic{algorithm}}
\renewcommand{\theHalgorithm}{S\arabic{algorithm}}

\section{SAT-based Gadget Search}
\label{S-sup:search}

\subsection{Search Region}

\begin{figure}[t]
\centering
\includegraphics{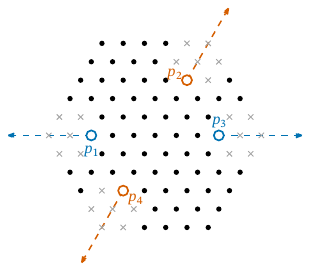}
\caption{\label{S-fig:sat-region}%
Search region for fixed pins. Dots and crosses: the region $W$. Each pin
$p_i$ (open circle) has an exit direction $d(p_i)$, and its line
$p_i+t\,d(p_i)$, $t\ge1$ (dashed arrow), must stay free. The two
channels are $p_1p_3$ and $p_2p_4$. Sites on a line or adjacent to one
(crosses) are excluded. The pins and the remaining sites (dots) form
$W_P$, the sites the solver may occupy.}
\end{figure}

For a set $S$ of lattice sites, $H_S$ is the graph it induces. The
\emph{line} of a pin $p$ is the ray of sites $p+t\,d(p)$, $t\ge1$,
reserved for its chain. Pin choices whose lines touch each other or
another pin are discarded. For the remaining choices, the
\emph{candidate sites} $W_P\subseteq W$ are the four pins together with
every site of $W$ that is neither on a line nor adjacent to one
(Fig.~\ref{S-fig:sat-region}). The pins are included although each is
adjacent to the first site of its own line.

\subsection{SAT Encoding}
\label{S-sec:encoding}

Fix the target $T=\rvalpha(R)$, the pins $(P,d)$, the atom count $N$ and
the offset $c$. A Boolean variable $x_v$ for each $v\in W_P$ states that
$v$ is occupied, so a satisfying assignment gives $S=\{v:x_v\}$. Each of
the four constraints of Sec.~\ref{sec:encoding} adds auxiliary
variables of its own.

\paragraph*{Atom placement}
Unit clauses occupy the pins, and $\sum_vx_v=N$ is encoded with
sequential counters~\cite{sinz2005}.

\paragraph*{Connectivity}
Variables $w^t_v$, $0\le t<N$, mark the sites that $H_S$ connects to
$p_1$ by a path of at most $t$ edges. Layer $t+1$ is defined from layer
$t$, and every occupied site must be marked in layer $N-1$.

\paragraph*{MIS lower bounds}
For each finite entry $T_{\mathbf s}$, a separate set of variables
selects a witness: an independent set of $H_S$ with boundary
configuration $\mathbf s$ and at least $T_{\mathbf s}+c$ vertices.

\paragraph*{MIS upper bounds}
An upper bound must hold for every independent set of $H_S$, so one
witness is not enough. We build a decision diagram that depends only on
$W_P$: it visits the candidate sites in row order and takes or skips
each one, and a site can be taken only if it has no taken neighbor. A
node records the current step and the taken sites that still have an
unvisited neighbor, which is all that later steps need. The paths of the
diagram are thus the independent sets of $H_{W_P}$. The occupancy
variables restrict them to $H_S$: a step that takes site $v$ is allowed
only when $x_v$ is true, so the remaining paths are the independent sets
of $H_S$. For each $\mathbf s$, variables mark the pairs of node and
count reached by these paths with configuration $\mathbf s$, and the end
of the diagram may not be reached with more than
$\max_{\mathbf t\preceq\mathbf s}T_{\mathbf t}+c$ vertices.

Because $T$ is reduced, the two MIS bounds hold together if and only if
$\rvalpha(H_S)=T+c$. Hence $\Phi$ is satisfiable if and only if some $S\subseteq W_P$ with
$P\subseteq S$ and $|S|=N$ induces a connected graph with
$\rvalpha(H_S)=T+c$.

\subsection{Search and Verification}
\label{S-sec:pin-selection}

Condition (G2) is enforced by three \emph{pin tests}, which are stricter
than (G2): each $d(p)$ points away from the centroid of the four first
sites; around that centroid, the first sites of $p_1,p_3$ alternate with
those of $p_2,p_4$; and each first site is an extreme point of the convex
hull of $S$ and the four first sites. Algorithm~\ref{S-alg:search}
increases $N$ and, for each $N$, enumerates the offsets and the pin
choices that satisfy (G1) and the first two tests, up to the lattice
symmetries that preserve $W$. It is therefore complete within $W$ and the
ranges of $N$ and $c$ for gadgets that pass the pin tests.

By Theorem~\ref{thm:reduction}, $H_S$ with boundary $P$ can replace
$R$ in every host graph, shifting the MIS size by $c$, exactly when
$\rvalpha(H_S)=\rvalpha(R)+c$. \textsc{IsGadget} checks this criterion
directly, independently of the SAT encoding: it computes the 16 values
$\valpha(H_S)_{\mathbf s}$ from $H_S$, reduces them, and compares the
result with $T+c$. It also tests connectivity, (G1) and the three pin
tests; the third depends on $S$ and is not in $\Phi$. A rejected $S$ is excluded by
the clause $\bigvee_{v\in S}\neg x_v$, which removes only $S$ because
$|S|=N$.

\begin{algorithm}[t]
\caption{SAT search}
\label{S-alg:search}
\begin{algorithmic}[1]
\Require target tensor $T$; region $W$; atom range $[N_{\min},N_{\max}]$;
  offset range $\mathcal C$
\Ensure gadget $(S,P,d,c)$ or \textsc{none}
\For{$N=N_{\min},\dots,N_{\max}$}
  \For{$c\in\mathcal C$ \textbf{and} pins $(P,d)$ in canonical order}
    \State $\Phi\gets\Phi(T,W_P,P,N,c)$ \Comment{Sec.~\ref{S-sec:encoding}}
    \While{$\Phi$ has a satisfying assignment $x$}
      \State $S\gets\{v:x_v=1\}$
      \If{\Call{IsGadget}{$S,P,d,T,c$}}
        \State \Return $(S,P,d,c)$
      \EndIf
      \State $\Phi\gets\Phi\land\bigvee_{v\in S}\neg x_v$ \Comment{block $S$}
    \EndWhile
  \EndFor
\EndFor
\State \Return \textsc{none}
\end{algorithmic}
\end{algorithm}

\section{Details of the Planarization Placement}
\label{S-sup:planar}

\subsection{Planarize}
\label{S-sup:planarize}

\paragraph*{Drawing}
A planar graph receives a straight-line grid
drawing~\cite{fary1948straight,defraysseix1990draw}. Otherwise
simulated annealing (Sec.~\ref{sec:planar}) runs for a fixed number of
sweeps from a force-directed layout~\cite{fruchterman1991graph} on an
integer grid and returns the drawing with the fewest crossings. Edges
that touch count as crossing.

\begin{figure}[t]
\centering
\includegraphics{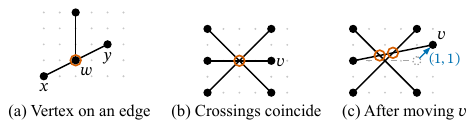}
\caption{\label{S-fig:degeneracy}%
Degeneracies of a grid drawing. (a) Vertex $w$ lies on edge $xy$, so the
chain along $xy$ would pass through $w$. (b) Three edges meet at one
point (circled), so the two crossings on each edge share a point, and no
crossing gadget, which joins two edges, fits there. (c) Moving $v$ by
$(k,k^2)$ with $k=1$ separates them into two crossings (circled), each
of which becomes one crossing gadget.}
\end{figure}

\paragraph*{General position}
The plane graph $\mathcal N$ turns every crossing into one crossing
gadget, which joins exactly two edges at a point inside both. Its
construction therefore needs a drawing in \emph{general position}: no
two vertices share a point, no vertex lies on an edge it is not an end
of, and no two crossings on an edge share a point. The annealed drawing
can violate this because its vertices lie on an integer grid, where
collinear points are common [Fig.~\ref{S-fig:degeneracy}(a,~b)]. Each such \emph{degeneracy} involves a set
of vertices: the two vertices, the vertex and the ends of the edge, or
the ends of the edge and of the two edges crossing it. Its largest
vertex is the largest of them in a fixed order. Planarize ends by
repeating the following until no degeneracy is left: let $v$ be the
smallest vertex that is the largest vertex of some degeneracy, and move
$v$ by $(k,k^2)$ for the smallest integer $k\ge1$ at which the drawing of
the subgraph induced by the vertices up to $v$ has no degeneracy
[Fig.~\ref{S-fig:degeneracy}(c)].

The positions of $v$ that create such a degeneracy lie on finitely many
points and lines. A line meets the parabola $\{v+(k,k^2)\}$ in at most
two points, so a suitable $k$ always exists. A move of $v$ creates no
degeneracy among smaller vertices, so the vertex chosen next is larger,
and at most $n$ moves are made.

\paragraph*{Plane graph}
The direct edge of $uv$ lies on the segment between crossings whose
midpoint is nearest the midpoint of $uv$. A copy with more than three
neighbors is split into copies with at most three, keeping the cyclic
order of its neighbors. Within each source vertex's COPY tree, we remove
unused leaf copies, retaining every endpoint of a source edge and at
least one copy per tree. We merge two adjacent copies of the same tree
if they have no common neighbor and the merged copy has at most three
neighbors.

\subsection{Shape}
\label{S-sec:planar-shape}

\paragraph*{Orthogonal drawing}
Tamassia's minimum-cost flow~\cite{tamassia1987} gives an orthogonal
drawing of $\mathcal N$ with the fewest bends, where a turn at a node of
degree two also counts as a bend, and OGDF~\cite{ogdf} assigns integer
coordinates.

\begin{figure}[t]
\centering
\includegraphics{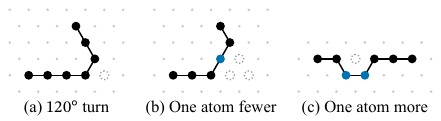}
\caption{\label{S-fig:parity}%
Chain shapes on the lattice. (a) At a $120^\circ$ turn the corner
(dotted) is left out, since its two chain neighbors are adjacent.
(b) Cutting one step deeper, one atom (blue) replaces the corner and its
two neighbors, so the chain has one atom fewer. (c) A detour replaces one
atom of a straight segment by two (blue), so the chain has one atom more.
In each case the chain stays an induced path.}
\end{figure}

\paragraph*{Lattice embedding}
The two drawing axes map to lattice directions $60^\circ$ apart, so each
bend becomes a turn of $60^\circ$ or $120^\circ$ in the direction of the
chain. At a $120^\circ$ turn the corner atom is left out, since its two
neighbors on the chain are adjacent [Fig.~\ref{S-fig:parity}(a)]. Each
node becomes a fixed set of atoms, and each chain leaves it as an arm
from a finite catalog. We checked by computer that every node with every
combination of arms induces exactly the intended edges.

\paragraph*{Separation}
Each distinct coordinate value of the drawing, in either axis, becomes a
band of the lattice that is wider than the nodes, arms and offsets placed
on it, and adjacent bands are at least four lattice units apart. Two
elements of the drawing that do not meet are separated by such a gap in
at least one axis, so their atoms are not adjacent. Every layout the
compiler returns is also checked to induce exactly the intended edges.

\begin{figure}[t]
\centering
\includegraphics{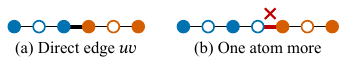}
\caption{\label{S-fig:path-parity}%
Copies (filled) and middle vertices (open) on a COPY gadget; atoms of $u$ are
blue and of $v$ orange. (a) A direct edge (thick) joins a copy of $u$
to a copy of $v$. (b) With one atom more, the direct edge would join a middle vertex of
$u$ to a copy of $v$.}
\end{figure}

\paragraph*{Parity}
Copies alternate with middle vertices along each COPY path, so a path
between two copies of the same source vertex has an even number of edges.
A direct edge must join a copy of $u$ to a copy of $v$
[Fig.~\ref{S-fig:path-parity}(a)]. Adding or removing one atom from a
connecting path flips its parity and can make the direct edge meet a
middle vertex [Fig.~\ref{S-fig:path-parity}(b)], breaking the encoding.

Moving a node can change the lengths of multiple paths, so their parity constraints may conflict. Shape first adjusts node positions and how paths leave them to satisfy a consistent subset of these constraints. The constraints form a linear system over $\mathrm{GF}(2)$, and we add them one at a time, keeping each only if the system remains solvable, as checked by Gaussian elimination.

Any remaining parity mismatch is corrected locally by changing the path's atom count by one (Fig.~\ref{S-fig:parity}). A path with a $120^\circ$ turn cuts that corner one step deeper, while other paths take a detour on a straight segment. Since paths with a $120^\circ$ turn can always be corrected locally, they are excluded from the linear system.

\subsection{Compact}
\label{S-sup:compact}

Shape spreads the drawing to keep its parts apart, so its paths are
longer than needed. Compact reduces the atom count by shortening them
while keeping three things: (1) the shape of the drawing, so every gadget
keeps its connections; (2) the parity of every path; and (3) the separation of
atoms that must not touch.

\begin{figure}[t]
\centering
\includegraphics{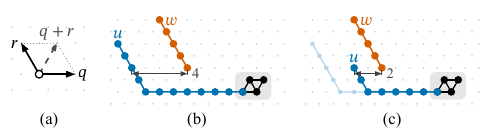}
\caption{\label{S-fig:compact}%
One compaction pass. (a) Coordinate $q$ increases along the horizontal
lattice direction and $r$ along the $120^\circ$ direction; a step along the
third direction (dashed), their sum, increases both. (b) Before the pass: the gadget (gray) moves rigidly, and the two segments of the path of
$u$ (blue) may shrink, as in set $S$. The paths of $u$ and $w$ (orange) must stay apart along
$q$, a pair in $C$ whose current gap is 4. (c) After the pass (faint:
before), each segment of $u$ is two atoms shorter, so its parity is kept,
and the gap has reached its bound of 2.}
\end{figure}

\paragraph*{Flow formulation}
In axial coordinates $(q,r)$, a step along one lattice direction changes
only $q$, a step along a second changes only $r$, and a step along the
third changes both [Fig.~\ref{S-fig:compact}(a)]. A pass optimizes $q$ and
$r$ separately. For $q$, path segments along the $q$ direction may
stretch or shrink, and everything else moves rigidly: a gadget and the
segments attached to it form a group $g$, shifted by $y_g$
[Fig.~\ref{S-fig:compact}(b,~c)]; $r$ is treated in the same way. A gadget
atom or segment end $e$ with coordinate $A(e)$ moves to
$x_e=A(e)+y_{g(e)}$, and the program is
\begin{equation}
\begin{aligned}
 \min\quad & \sum_{(e,f)\in S}\bigl(x_f-x_e\bigr)\\
 \text{s.t.}\quad & x_f-x_e\ge b_{ef},\quad (e,f)\in S\cup C.
\end{aligned}
 \label{S-eq:lp}
\end{equation}
Here $S$ contains the stretchable segments, from lower end $e$ to upper
end $f$, with $b_{ef}=1$, and $C$ the pairs of atoms that must not touch,
with $b_{ef}\in\{1,2\}$, each kept on its current side. Every constraint
is a difference of two variables, so \eqref{S-eq:lp} is the dual of a
minimum-cost flow and has an integral optimum~\cite{schrijver1986}.

\paragraph*{Third direction}
A segment along the third direction changes $q$ and $r$ at once, so
neither program of the pass can shorten it, and \eqref{S-eq:lp} keeps
its length fixed. To shorten such segments as well, the pass first
solves a relaxation in which all coordinates move together. It proposes
new lengths for these segments, rounded to keep their parity, which
\eqref{S-eq:lp} then uses as fixed lengths. Since \eqref{S-eq:lp} still
enforces every constraint, rounding cannot break one; if the proposed
lengths cannot be met, the old ones are kept.

\paragraph*{Parity and contacts}
A path changes parity only if the groups at its two ends shift by
amounts of different parity. Each pass therefore fixes a parity $p_g$
for every group that keeps all path parities and restricts the shifts to
$y_g=2m_g+p_g$, which keeps \eqref{S-eq:lp} a flow dual. Pairs in $C$ are
added lazily: when a solution creates an unintended contact, that pair
joins $C$ and the pass is solved again. Passes rotate the roles of the
three lattice directions, so each in turn is the third, until none
shortens the layout.

\section{Simulation Setup}
\label{S-sup:setup}

\subsection{Source Graph}
\label{S-sup:large-graphs}

Fig.~\ref{fig:k33-layouts} shows the source graph $K_{3,3}$ and its
two layouts. The triangular layout comes from planarization placement
and has 79 atoms, MIS size 39 and two maximum independent sets. The King's
layout comes from \texttt{pred}~\cite{pan2026problem}, keeping the
smallest layout over a search of vertex orders and compaction. It has 50
atoms, MIS size 23 and two maximum independent sets.

\subsection{Atom Spacing}
\label{S-sup:endpoint}

The simulations neglect decoherence, atom loss and position errors. Both
lattices set their spacing by the rule of the Bloqade.jl
documentation~\cite{bloqade_blockade},
$\Omega_{\max}=\sqrt{V(r_{\max})V(r_{\min})}$, where $r_{\max}$ is the
longest edge and $r_{\min}$ the shortest non-edge distance
(Table~\ref{S-tab:hw}). This places \(\Omega_{\max}\) geometrically between the weakest edge and strongest non-edge interactions, balancing their separation from the drive scale. On
the King's lattice it gives $R_b/a=1.68$, close to the value $1.7$ used
in experiment~\cite{ebadi2022quantum}.

\begin{table}[h]
\centering
\caption{\label{S-tab:hw}Atom spacing and interaction strengths. Interactions in units
of $\Omega_{\max}$.}
\begin{tabular}{lcc}
\toprule
Quantity & Triangular & King's \\
\midrule
Spacing $a$ ($\mu$m) & 5.88 & 4.60 \\
Weakest edge $V(r_{\max})$ & 5.20 & 2.83 \\
Other edges & --- & 22.6 \\
Strongest non-edge $V(r_{\min})$ & 0.19 & 0.35 \\
\bottomrule
\end{tabular}
\end{table}

\subsection{Numerical Simulation}
TDVP takes two-site steps with third-order global subspace expansion
until the bond dimension reaches $\chi=64$, and one-site steps after. The
time step is $\delta t=0.1/\Omega_{\max}\approx4.0$~ns, shortened only
where a step would pass a corner of the schedule. Each simulation draws
$10^4$ samples; sampling uncertainty is shown as 95\% Wilson intervals
in Fig.~\ref{fig:large-dyn}.

\end{document}